\documentclass[aps,prc,reprint,superscriptaddress,nofootinbib,nolongbibliography]{revtex4-2}

\usepackage[utf8]{inputenc}
\usepackage{CJK}
\usepackage{hyperref,graphicx,amsmath,braket}

\begin{document}
\begin{CJK*}{UTF8}{gbsn}

\title{Evidence for shape coexistence in $^{120}$Sn from\\ 
the first $0^+_3$ lifetime measurement}

\author{F.~Wu (吴桐安)} \email[]{twa73@sfu.ca}
\author{C.~Andreoiu} 
\affiliation{Department of  Chemistry, Simon Fraser University, Burnaby, BC, V5A 1S6, Canada}
\author{V.~Karayonchev}
\affiliation{Argonne National Laboratory, Argonne, IL, 60439, USA}
\author{C.M.~Petrache}
\affiliation{Universit\'e Paris-Saclay, CNRS/IN2P3, IJCLab, Orsay, 91405, France}
\author{J.-M.~R\'egis}
\affiliation{Institut f\"ur Kernphysik, Universit\"at zu K\"oln, K\"oln, 91405, Germany}
\author{A.~Esmaylzadeh}
\affiliation{Institut f\"ur Kernphysik, Universit\"at zu K\"oln, K\"oln, 91405, Germany}
\author{C.~Michelagnoli}
\affiliation{Institut Laue-Langevin, Grenoble, F-38042, France}
\author{M.~Beuschlein}
\affiliation{Technische Universit\"at Darmstadt, Department of Physics,
Institute for Nuclear Physics, 64289 Darmstadt, Germany}
\author{P.~Spagnoletti}
\affiliation{Department of  Chemistry, Simon Fraser University, Burnaby, BC, V5A 1S6, Canada}

\author{G.~Colombi}
\author{J.M.~Daugas}
\author{L.~Domenichetti}
\affiliation{Institut Laue-Langevin, Grenoble, F-38042, France}
\author{P.E.~Garrett}
\affiliation{Department of Physics, University of Guelph, Guelph, ON, N1G 2W1, Canada}
\affiliation{Department of Physics, University of the Western Cape, P/B X17, Bellville ZA-7535, South Africa}
\author{J.~Jolie}
\author{M.~Ley}
\affiliation{Institut f\"ur Kernphysik, Universit\"at zu K\"oln, K\"oln, 91405, Germany}
\author{S.~Pannu}
\affiliation{Department of Physics, University of Guelph, Guelph, ON, N1G 2W1, Canada}
\author{E.~Taddei}
\affiliation{Department of  Chemistry, Simon Fraser University, Burnaby, BC, V5A 1S6, Canada}

\date{May 7, 2025}

\begin{abstract}
The lifetime of the $0^+_3$ state in $^{120}$Sn was measured for the first time applying the fast-timing technique following thermal neutron capture. The mean lifetime of $\tau = 50(7)$~ps leads to a $E0$ transition strength of $10^3\times \rho^2(E0;0^+_3\rightarrow0^+_2) = 120(50)$, suggesting shape coexistence and a high degree of mixing between the $0^+_2$ and $0^+_3$ states. With the newly measured lifetime, the $B(E2;0^+_3\rightarrow 2^+_1)$ value is 0.50(7)~W.u., which reveals that the $\rho^2(E0;0^+_3\rightarrow0^+_1)$ increases by a factor of $\approx 3.4$ from $^{116}$Sn to $^{120}$Sn.
\end{abstract}
\maketitle
\end{CJK*} 

The semi-magic $Z=50$ Sn isotopes, which possess the largest number of stable isotopes of any element, constitute one of the best studied isotopic chains in the nuclear chart. The neutron mid-shell $^{114-122}$Sn nuclei have been of special interest because they exhibit shape coexistence, where the excited $0^+$ states built on proton 2 particle-2 hole~(2p-2h) configurations intrude into the low-lying states with two-neutron configurations~\cite{BRON1979335,PhysRevC.99.024303,Paul_SC,Leoni2024}. However, the level energies alone are insufficient to firmly establish the nature of these $0^+$ states; detailed spectroscopic investigations of the electromagnetic transition rates are required in order to determine the collectivity and the degree of mixing of these states~\cite{Leoni2024}. For example, systematic investigations of $B(E2;0^+_1\rightarrow 2^+_1)$ values~\cite{PhysRevC.110.054304,PhysRevC.99.034609,PhysRevC.97.054319,PhysRevC.96.054318,PhysRevC.92.041303,PhysRevC.84.061303,Jungclaus2011}, supported by Monte Carlo Shell Model~(MCSM) interpretations~\cite{PhysRevLett.121.062501}, suggest that the $0^+_\text{g.s.}$ and the $2^+_1$ states of the Sn isotopes could be oblate deformed~\cite{Leoni2024,PhysRevLett.121.062501}. Proton excitations across the $Z=50$ shell gap were required to reproduce the observed $B(E2)$ strengths in the chain of Sn isotopes~\cite{PhysRevLett.121.062501}, which deviate from the simple parabola predicted by the exact neutron seniority scheme~\cite{Morales2011,PhysRevC.72.061305}.\par

The $E0$ and $E2$ transitions from the excited $0^+$ states in the even-even $^{112-124}$Sn isotopes have been systematically studied with comprehensive $\gamma$-ray and electron spectroscopy employing a rich assortment of coincidence techniques by the collaboration of B\"acklin {\it et al.}~\cite{Bcklin1981,Jonsson1981,Kantele1979} using different population mechanisms such as Coulomb excitation, ($p$, $p'$), and $\beta$ decay. A strong $E0$ transition with $\rho^2(E0;0^+_3\rightarrow0^+_2)$ = 100(20) was observed in $^{116}$Sn~\cite{Bcklin1981,Kantele1979}, suggesting that the $0^+_2$ and $0^+_3$ states are strongly mixed and have different deformations~\cite{Kantele1979}. For $^{120}$Sn, only upper limits for the absolute $E0$ and $E2$ transition rates from the $0^+_3$ state could be determined from Coulomb excitation~\cite{Bcklin1981}. However, the branching ratios measured in Ref.~\cite{Bcklin1981} indicate a strong $E0(0^+_3\rightarrow0^+_2)$ transition also in $^{120}$Sn. From the branching ratios, the ratio $\rho^2(E0;0^+_3\rightarrow0^+_2) / \rho^2(E0; 0^+_3\rightarrow0^+_1) = 36(16)$ and the dimensionless $X(E0/E2)$ ratio of $B(E0;0^+_3\rightarrow0^+_1) / B(E2; 0^+_3\rightarrow2^+_1) = 0.22(5)$ were extracted~\cite{Bcklin1981}. For $^{116}$Sn, these ratios are 110(10) and 0.066(11), respectively~\cite{Bcklin1981}. Therefore, a lifetime measurement of the $0^+_3$ state in $^{120}$Sn is required to determine the associated absolute $E0$ and $E2$ transition rates and thus the origin of the differences of these ratios. 

More recently, experimental investigations of the transition rates in $^{116}$Sn showed that the $0^+_3$ state was more deformed than the $0^+_2$ state, and therefore was assigned as the band-head of the 2p-2h rotational band~\cite{Pore2017}, while Interacting Boson Model~(IBM-2) calculations showed strong mixing between the $0^+_2$ and $0^+_3$ states~\cite{PhysRevC.99.024303}. In $^{118}$Sn, the $0^+_2$ state is considered to be the head of the deformed intruder band~\cite{PhysRevC.102.024323} with evidence for strong mixing with the $0^+_3$ state with $10^3 \times \rho^2(E0;0^+_3\rightarrow0^+_2)>38$~\cite{PhysRevC.109.054317}. To support these conclusions, transition rates in other Sn isotopes are necessary, in particular in the neighboring $^{120}$Sn nucleus. \par
 For $^{120}$Sn, upper limits of the $\rho^2(E0)$ values from the $0^+_3$ state in the literature evaluations and reviews~\cite{Kibdi2005,KIBEDI2022103930,Paul_SC,Leoni2024} differ by a factor of $\approx$ 150. The 2005 evaluation~\cite{Kibdi2005} reports values of $\rho^2(E0;0^+_3\rightarrow0^+_2)<1000$ and $\rho^2(E0;0^+_3\rightarrow0^+_1) < 27$, while the 2022 evaluation~\cite{KIBEDI2022103930} reports values $<7$ and $<0.17$, respectively. The differences likely stemmed from the treatment of the  lower limit of the half-life $T_{1/2} > 4$~ps~\cite{Bcklin1981}. Given the large method-dependent discrepancy between the evaluations~\footnote{In the 2005 evaluation~\cite{Kibdi2005}, the $\rho^2(E0)$ values were calculated at the lower limit of $T_{1/2} = 4$~ps. In the 2022 evaluation~\cite{KIBEDI2022103930}, a more sophisticated Monte-Carlo error analysis, treating the probability density function~(PDF) of the $0^+_3$ half-life as a uniform distribution from the lower limit of 4~ps to 5.8~ns~\cite{kibedi_email} was used to determine the PDF of the $\rho^2(E0)$ values.}, a direct lifetime measurement of the $^{120}$Sn $0^+_3$ state is required to shed more light on the degree of mixing of the $0^+$ coexisting configurations in $^{120}$Sn.\par

\begin{figure}[ht]
\centering
\includegraphics[width=\linewidth]{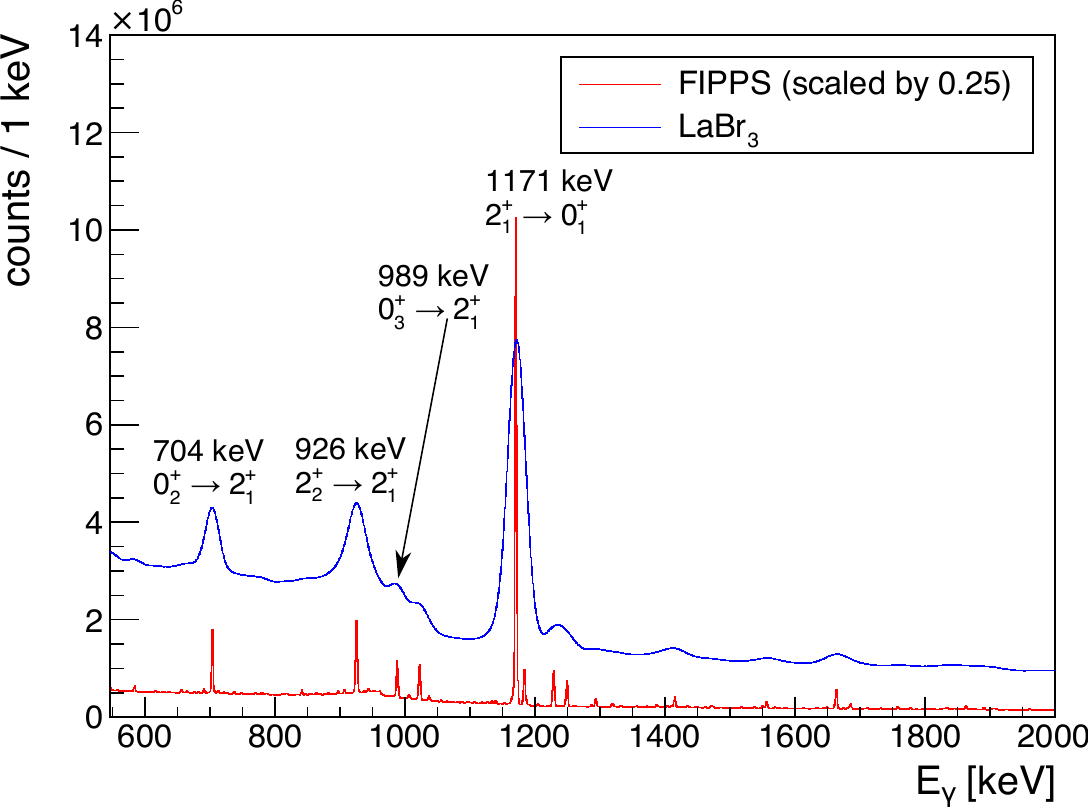}
\caption{Overlapping total projections from the $\gamma\gamma\gamma$ cube for the FIPPS HPGe~(red) and LaBr$_3$~(blue) detectors. The labels on the peaks correspond to transitions in $^{120}$Sn.}\label{fig-proj}
\end{figure}

In this work, the $0^+_3$ state in $^{120}$Sn was populated using a thermal neutron capture reaction on a 83.98\% enriched $^{119}$Sn target at the Institut Laue-Langevin~(ILL) in Grenoble, France. The excited $^{120}$Sn nucleus was formed at the neutron separation energy, $S_n = 9.1$~MeV. It then decayed to the ground state via the emission of $\gamma$ rays, which were detected using the FIssion Product Prompt gamma-ray Spectrometer (FIPPS)~\cite{FIPPS}, consisting of eight clover-type HPGe detectors with anti-Compton BGO shielding, coupled to an array of 15 LaBr$_3$ fast-timing detectors, similar to that in Ref.~\cite{PhysRevC.99.024303}. Experimental details and the data sorting procedure are reported in Ref.~\cite{WU2025123105}. The maximum neutron flux at the FIPPS experimental station was 10$^8$~s$^{-1}$cm$^{-2}$, and the cross section for the $^{119}$Sn(n,$\gamma$)$^{120}$Sn reaction is 2.2(5)~b~\cite{etde_20332542}. After 14 days of beam on target, 4.3$\times 10^9$ $\gamma$$\gamma$$\gamma$ events were recorded~\cite{ILL_Data_120Sn} consisting of FIPPS-LaBr$_3$-LaBr$_3$ coincidences, after excluding signals in neighboring LaBr$_3$ detectors for active Compton suppression. The projections from the $\gamma$$\gamma$$\gamma$ cube are shown in Fig.~\ref{fig-proj}. Nearly all visible peaks correspond to transitions in $^{120}$Sn, indicating that there was little contamination from isotopic impurities. \par
\begin{figure}[h!]
    \centering
    \includegraphics[width=\linewidth]{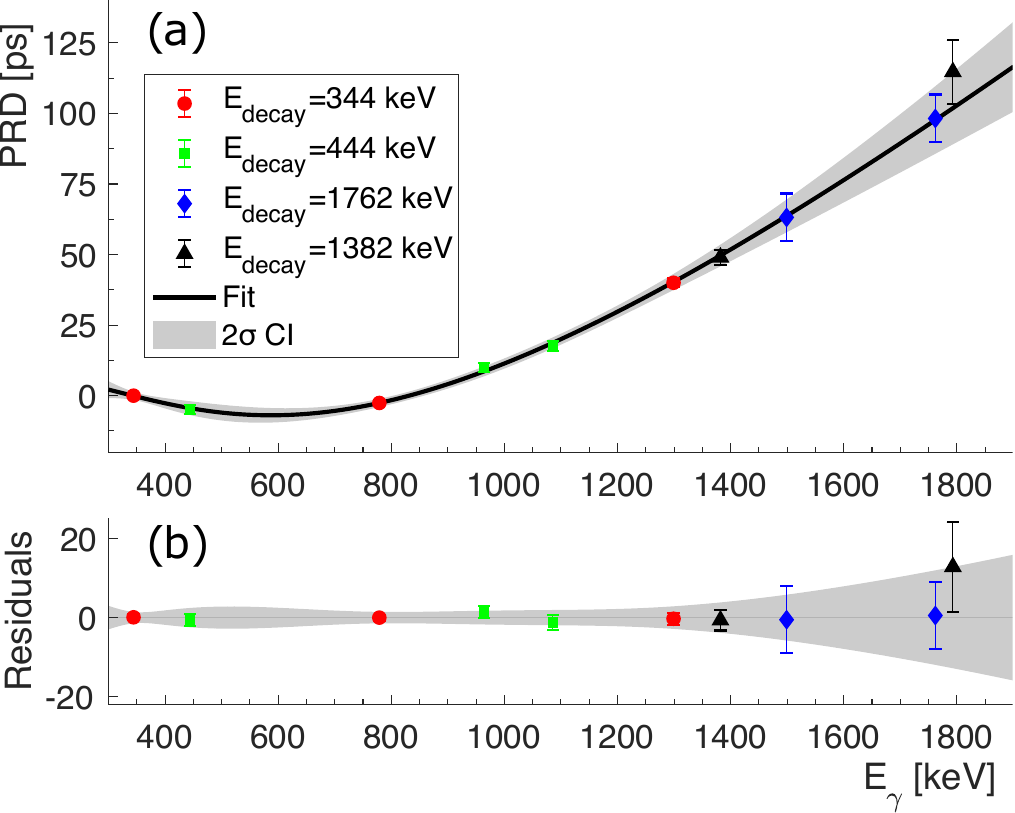}
    \caption{(a) The PRD fitted using Eq.~\ref{eq:PRD_fit_VK} and (b) the obtained residuals for the FIPPS-LaBr$_3$ setup that was used for the $^{120}$Sn lifetime measurement. The red circles and green squares correspond to transitions from a $^{152}$Eu standard source, while the blue diamonds and black triangles correspond to transitions from the $^{48}$Ti(n,$\gamma$)$^{49}$Ti calibration reaction. The gray band represents the $2\sigma$ confidence interval (CI) from the fit.} 
    \label{fig:prd-fit}
\end{figure}

The lifetime of the $0^+_3$ state was extracted using the fast-timing technique with the Generalized Centroid Difference~(CGD) method~\cite{REGIS201672,REGIS2025104152}, where the time difference between the two LaBr$_3$ signals populating and depopulating the $0^+_3$ state was measured using time-to-amplitude converters (TACs). The TAC events were then correlated offline to the energies recorded by the two LaBr$_3$ detectors, and a ``delayed'' time distribution was generated if the TAC was started by the feeding transition and stopped by the decay transition, and an ``anti-delayed'' time distribution if vise-versa. The lifetime of the state, without the presence of time-correlated background, can then be related to the centroid difference of the two distributions by
\begin{equation}
\label{eq:gcd}
    \Delta C = 2\tau + {\text{PRD}}(E_{\text{feeder}}, E_{\text{decay}})~,
\end{equation}
where $\Delta C$ is the centroid difference between the delayed and anti-delayed TAC distributions, and the prompt-response difference, ${\text{PRD}}(E_{\text{feeder}}, E_{\text{decay}})$, is the energy-dependent time walk of the fast-timing system. The PRD in this experiment was established using the $\Delta C$ from transitions feeding and decaying from states with known lifetimes with a $^{152}$Eu source and the $^{48}$Ti(n,$\gamma$)$^{49}$Ti reaction. The PRD at each $\gamma$-ray energy, $E_\gamma$, was calculated using Eq.~\ref{eq:gcd}, and fitted with
\begin{equation}
\label{eq:PRD_fit_VK}
    PRD = \cfrac{A}{\sqrt{E^2_\gamma + B}} + CE_\gamma + D~,
\end{equation}
where $A$-$D$ are free parameters, as described in Ref~\cite{VasilPhysRevC.99.024326}. The PRD with the best-fit is shown in Fig.~\ref{fig:prd-fit}. 

\begin{figure}[h!]
    \centering
    \includegraphics[width=\linewidth]{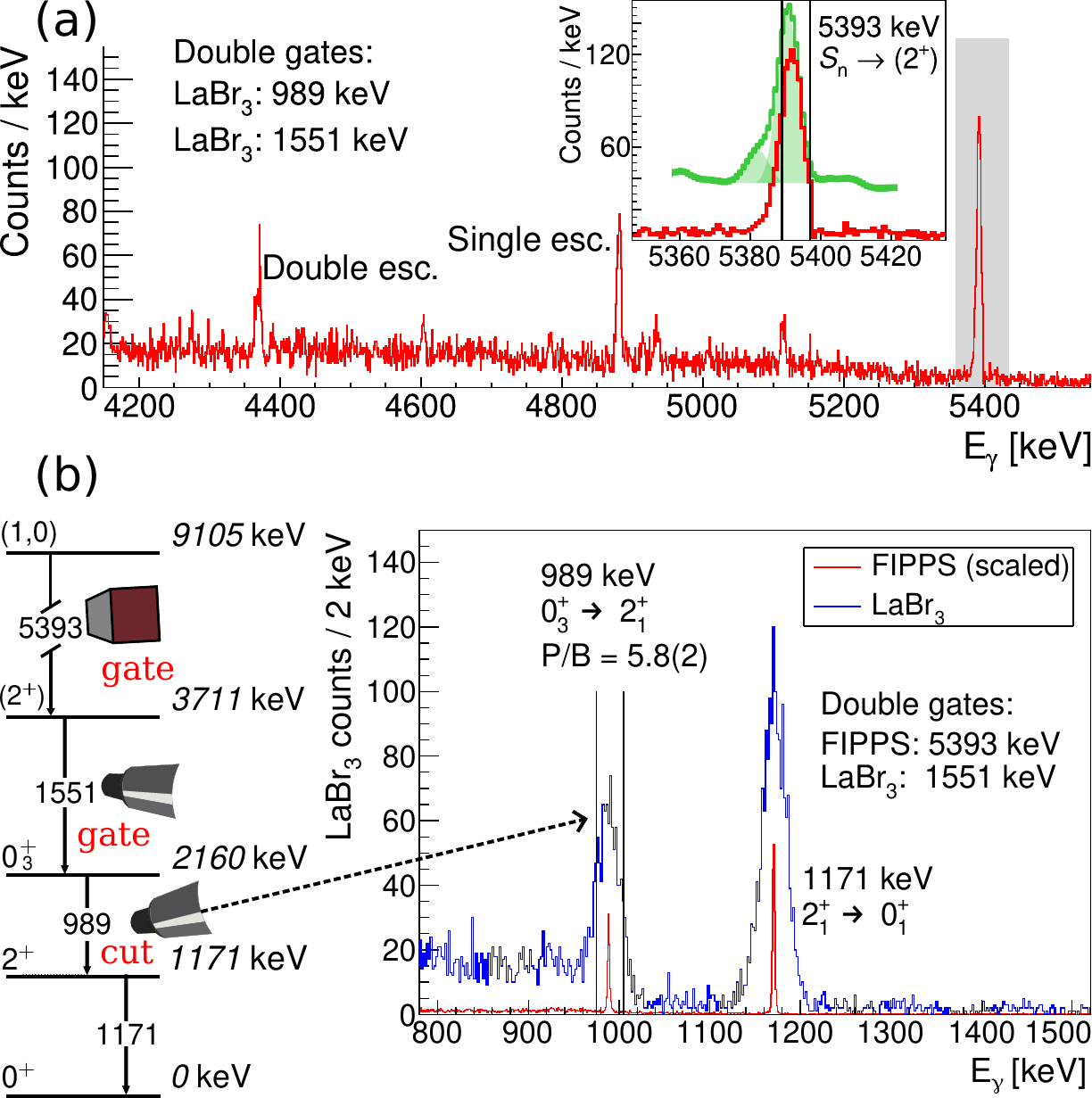}
\caption{(a): The FIPPS-HPGe spectrum from FIPPS-LaBr$_3$-LaBr$_3$ coincidence events with two LaBr$_3$ gates at 989~keV and 1551~keV. Inset: The FIPPS spectrum expanded around the shaded photopeak of the 5393-keV primary transition. The 5393-keV gate, illustrated by the vertical bars, is used for subsequent fast-timing analyses. The gate is off-centre in order to avoid the tail of a neighbouring peak in the un-gated FIPPS-HPGe spectrum on the left of the 5393-keV peak. The HPGe-singles spectrum (scaled) and the fit decomposition are shown in green in the inset. The escape peaks are not used for the analysis due to their lower peak-to-background ratios. Different HPGe gates, such as around the full photopeak and summing the photopeak with the single-escape peak were investigated and had no statistically significant impact on the final lifetime. (b) Left: Partial level scheme showing the population and decay of the $0^+_3$ state in $^{120}$Sn. (b) Right: the LaBr$_3$ spectrum~(in blue) from FIPPS-LaBr$_3$-LaBr$_3$ coincidence events and FIPPS-HPGe spectrum~(in red) from FIPPS-FIPPS-LaBr$_3$ coincidence events with a FIPPS-HPGe gate on the primary 5393-keV transition and a LaBr$_3$ gate on 1551-keV feeder transition.}
    \label{fig:0+3_showing_decay}
\end{figure}

\begin{figure}[h!]
    \centering
    \includegraphics[width=\linewidth]{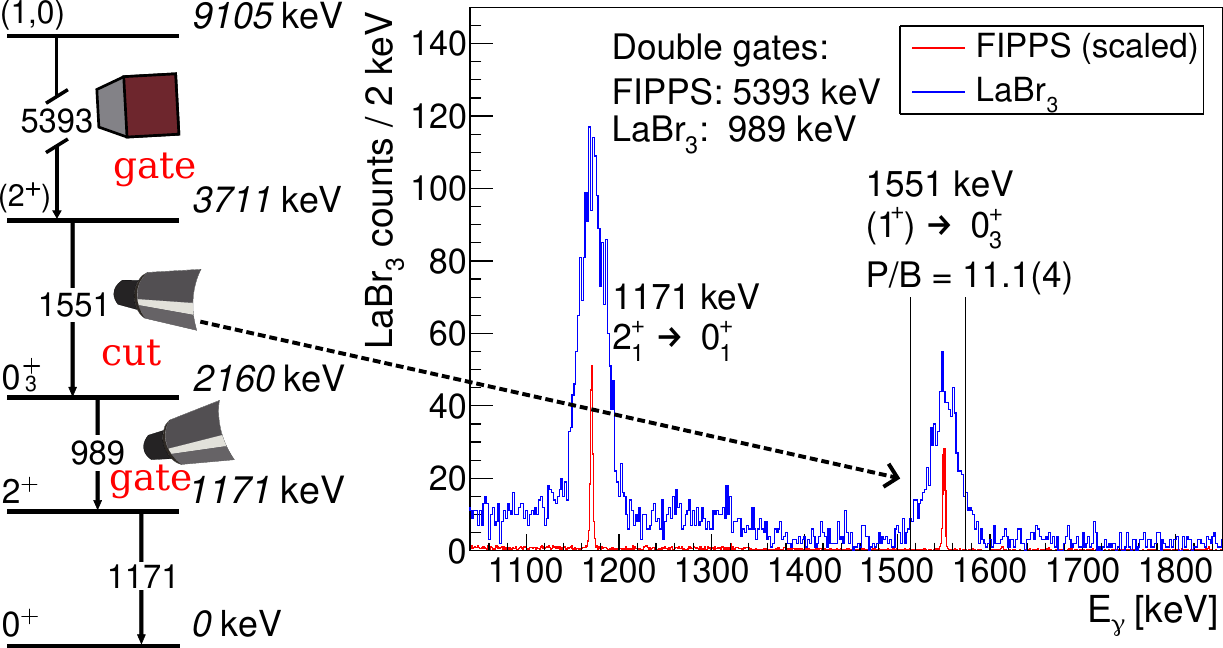}
\caption{Left: Partial level scheme showing the population and decay of the $0^+_3$ state in $^{120}$Sn. Right: the LaBr$_3$ spectrum~(in blue) from FIPPS-LaBr$_3$-LaBr$_3$ coincidence events and FIPPS-HPGe spectrum~(in red) from FIPPS-FIPPS-LaBr$_3$ coincidence events with a FIPPS-HPGe gate on the primary 5393-keV transition and a LaBr$_3$ gate on 989-keV decay transition. }
    \label{fig:0+3_showing_feeder}
\end{figure}

\begin{figure}[h!]
    \centering
    \includegraphics[width=\linewidth]{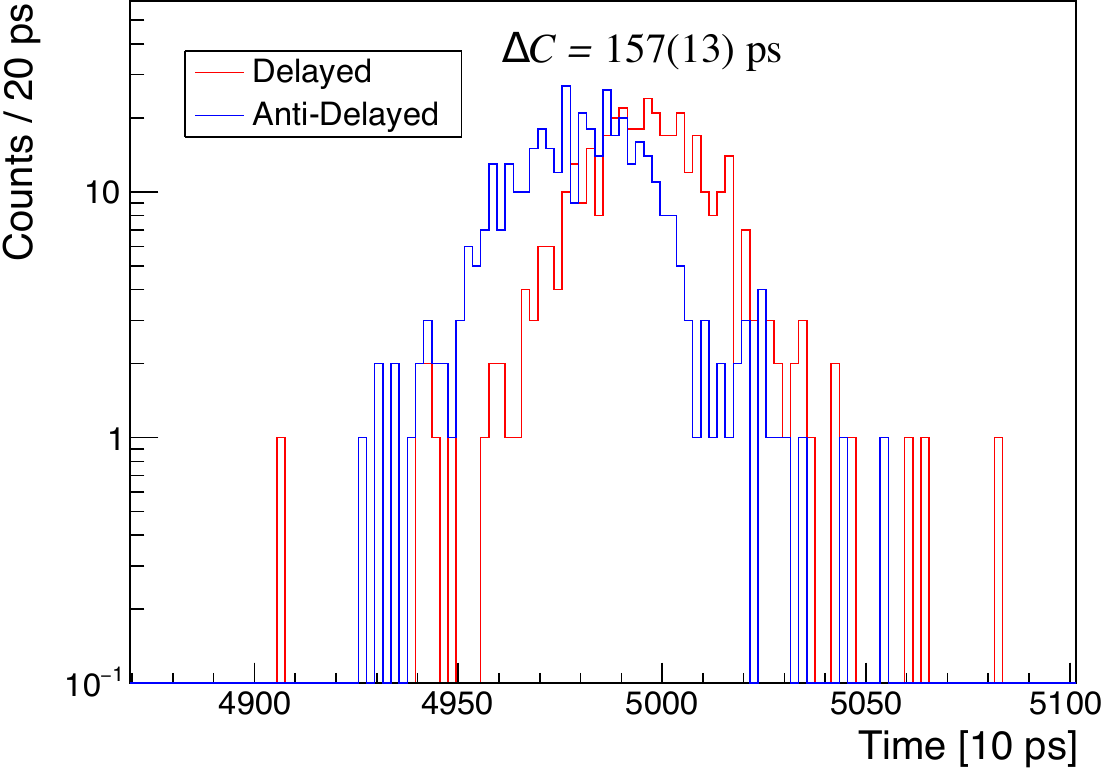}
\caption{The delayed and anti-delayed TAC spectra from the LaBr$_3$ gates, shown as vertical bars in Figs.~\ref{fig:0+3_showing_decay} and~\ref{fig:0+3_showing_feeder}, and the FIPPS-HPGe gate on the primary 5393-keV transition.}
    \label{fig:TAC_0+3}
\end{figure}

In the FIPPS-LaBr$_3$-LaBr$_3$ $\gamma$$\gamma$$\gamma$ cube, the high-resolution FIPPS-HPGe peak was used to precisely select the population of the ``feeder'' state at 3711~keV by gating on the primary 5393-keV transition, as illustrated in Fig.~\ref{fig:0+3_showing_decay}, and the two LaBr$_3$ peaks were used to obtain the TAC signal for lifetime extraction. Different HPGe gates, such as around the full photopeak and summing the photopeak with the single-escape peak were investigated and had no statistically significant impact on the final lifetime. The projection of the $\gamma$$\gamma$$\gamma$ cube on one LaBr$_3$, with a FIPPS gate on the 5393-keV primary transition and a LaBr$_3$ gate on the 1551-keV ``feeding'' transition is shown on the right of Fig.~\ref{fig:0+3_showing_decay}~(b). 
The overlaid red spectrum is the projection from FIPPS-FIPPS-LaBr$_3$ triple-coincident events under the same gates, showing that no additional transitions are present that are obscured by the widths of the LaBr$_3$ peaks. The spectra are nearly background free, and the only two peaks observed correspond to the transitions shown in the partial level scheme. Similarly, the projections with the same FIPPS gate, but with the LaBr$_3$ gate on the 989-keV ``decay'' transition are shown in Fig~\ref{fig:0+3_showing_feeder}. The spectra are also nearly background-free, where the only two peaks observed correspond to transitions in the partial level scheme. These spectra indicate that the TAC events correlated to these coincidence conditions indeed came from the feeder and decay transitions of 1551 and 989~keV, and that the difference in centroids of the delayed and anti-delayed TAC spectra are due solely to the lifetime of the $0^+_3$ state. The background contribution to the TAC spectra was investigated, and the corrections are below statistical uncertainties.
The TAC spectra, in coincidence with the 5393-keV primary transition in a FIPPS HPGe and the 1551-keV feeder and 989-keV decay transitions in the two LaBr$_3$ detectors, are shown in Fig.~\ref{fig:TAC_0+3}. From these TAC spectra we obtain a centroid difference of $\Delta C = 157(13)$~ps.\par
Substituting the centroid difference of 157(13)~ps and $PRD$(1551,989) of 59.6(71)~ps into Eq.~\ref{eq:gcd}, the $0^+_3$ lifetime was determined to be 50(7)~ps. \par
Using the same procedure, the time differences between the transitions feeding into and decaying from the 1171~keV $2^+_1$, 1875~keV $0^+_2$, and 2097~keV $2^+_2$ states were also measured. However, we were only able to establish upper limits on these lifetimes due to the limitation in sensitivity. These lifetimes are summarized in Tab.~\ref{tab:lifetimes}. 
\begin{table}[h!]
\caption{\label{tab:lifetimes}%
The lifetimes measured in this work compared to literature values~\cite{NDS_120}. 2$\sigma$ upper limits are reported for lifetimes shorter than the sensitivity of the setup.}
\begin{ruledtabular}
\begin{tabular}{llll}
\textrm{$E_\text{level}$ (keV)}&
\textrm{$J^\pi$}&
\textrm{$\tau_\text{this work}$~(ps)}&
\textrm{$\tau_\text{NNDC}$~(ps)}\\
\colrule
1171 & 2$^+_1$ &  $< 11$ & $0.92(17)$\\
1875 & $0^+_2$ & $ < 18$ & $10.7(14)$\\
2097 & $2^+_2$ &  $ < 13$ & $1.9(6)$ \\
2160 & $0^+_3$ & $50(7)$ & $>6$
\end{tabular}
\end{ruledtabular}
\end{table}

\begin{figure}[ht]
\centering
\includegraphics[width=\linewidth]{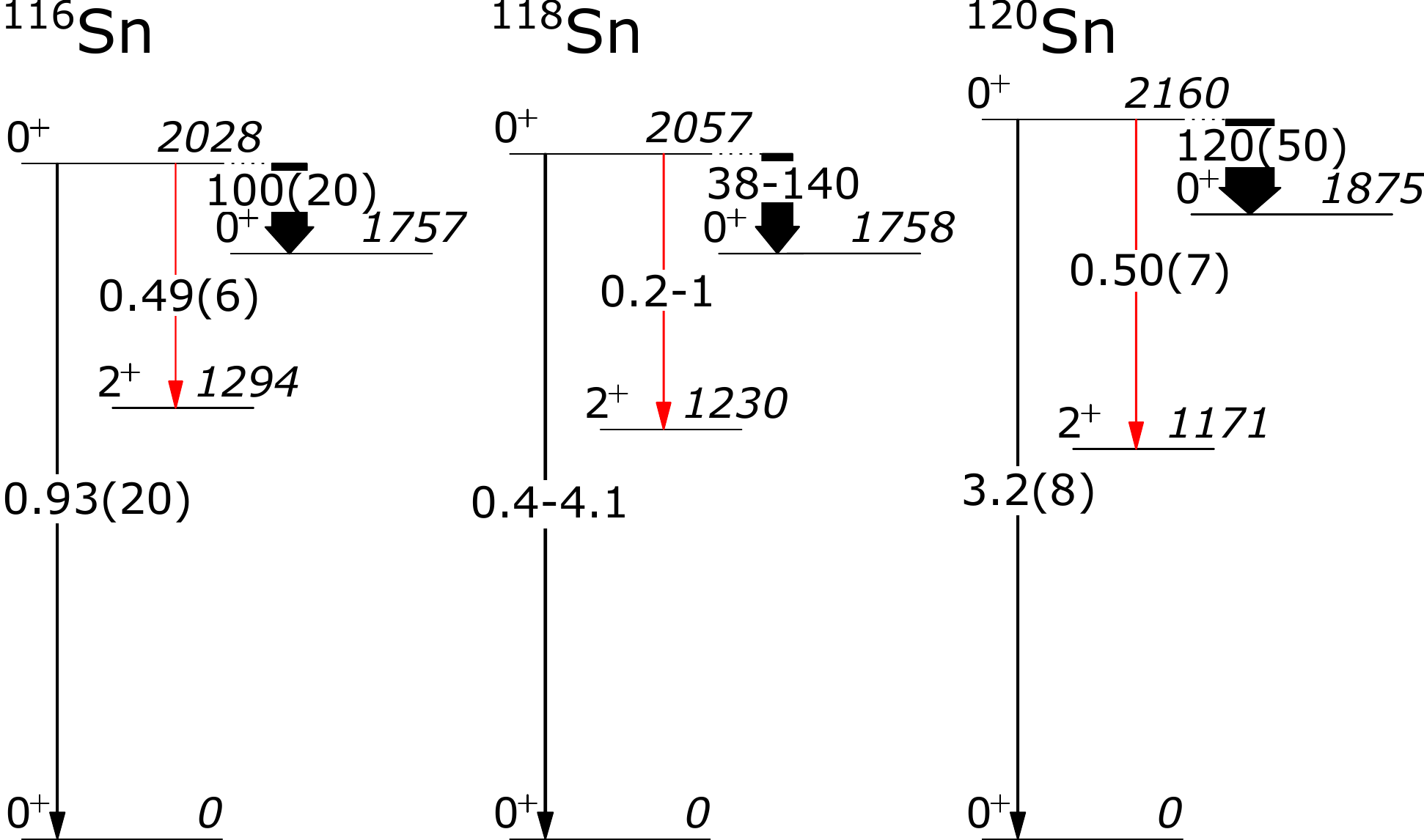}
\caption{The electromagnetic transition rates from the $0^+_3$ states in $^{116,118,120}$Sn. The $B(E2)$ values~(in red) are in W.u., and the $10^3\rho^2(E0)$ values are in black. The widths of the arrows are proportional to the transition strengths. The values for $^{116,118}$Sn are taken from Ref.~\cite{Bcklin1981} and Ref.~\cite{PhysRevC.109.054317}. The values for $^{120}$Sn are from this work. Note that the upper limits of the $B(E2; 0^+_3\rightarrow2^+_1)$ and, consequently, the $\rho^2(E0;0^+_{3}\rightarrow0^+_{1,2})$ values in $^{118}$Sn were extracted using Coulomb excitation by B\"acklin {\it et al.}~\cite{Bcklin1981,Jonsson1981,UUIP999,UUIP1001}, but they are not included in recent evaluations~\cite{NDS_120,Kibdi2005,KIBEDI2022103930,Paul_SC,Leoni2024} likely due to the formatting of Table 4 in Ref.~\cite{Bcklin1981}.}
\label{fig:em_rates}      
\end{figure}

The transition strengths from the $0^+_{3}$ state, updated with the newly measured $0^+_3$ lifetime of 50(7)~ps, are summarized in Tab.~\ref{tab:transitions} and the right of Fig.~\ref{fig:em_rates}. The $\rho^2(E0)$ values were calculated using the evaluated branching ratios from Ref.~\cite{KIBEDI2022103930,Bcklin1981}, and their uncertainties were determined through both conventional error propagation by adding the uncertainty contributions in quadrature and with the fully Monte-Carlo approach~\cite{KIBEDI2022103930}. The $\rho^2(E0)$ uncertainties, calculated with both methods, are in agreement. Our results are consistent with the upper limits of $E0$ values from Ref.~\cite{Bcklin1981} and the 2005 $E0$ evaluation~\cite{Kibdi2005}, but are $\approx$17 times larger than those from the 2022 evaluation~\cite{KIBEDI2022103930}, which have been used in other recent review articles such as Refs.~\cite{Leoni2024,Jenkins2023}. \par

\begin{table}[ht!]
\caption{\label{tab:transitions}%
The transition strengths from the $0^+_3$ state in $^{120}$Sn calculated using $\tau$ = 50(7)~ps of this work compared to literature values. The uncertainties in the $10^3\rho^2(E0)_\text{exp}$ values were determined using the fully Monte-Carlo approach as described in Ref.~\cite{KIBEDI2022103930}.}
\begin{ruledtabular}
\begin{tabular}{llll}
\textrm{$J^\pi_i \rightarrow J^\pi_f$}&
\textrm{$B(E2)_\text{exp}$ (W.u.)}&
\textrm{$10^3\rho^2(E0)_\text{exp}$}&
\textrm{$10^3\rho^2(E0)_\text{lit.}$}\\
\colrule
$0^+_3\rightarrow 0^+_1$ & - & 3.2$^{+0.9}_{-0.7}$ & $<0.17$~\cite{KIBEDI2022103930}\\
&  &  & $<27$~\cite{Kibdi2005}\\
&  &  & $<30$~\cite{Bcklin1981}\\
$0^+_3\rightarrow 0^+_2$ & - & 120(50) & $<7$~\cite{KIBEDI2022103930}\\
&  & & $<1000$~\cite{Kibdi2005}\\
&  & & $<1300$~\cite{Bcklin1981}\\
$0^+_3\rightarrow 2^+_1$ & 0.50(7) &  - & \\
\end{tabular}
\end{ruledtabular}
\end{table}

According to the simple relationship for $\rho(E0)$~\cite{KIBEDI2022103930,WOOD1999323}, 
\begin{equation}
    \label{eq:rho2e0}
    \rho(E0) \simeq \alpha \beta \Delta\braket{r^2},
    \end{equation}
where $\alpha$ and $\beta$ are the mixing coefficients and $\Delta\braket{r^2}$ is the difference in mean-squared charge radii of the two states, the large $10^3\rho^2(E0;0^+_3\rightarrow 0^+_2)$ value of 120(50) observed in our experiment suggests that the 0$^+_2$ and 0$^+_3$ states in $^{120}$Sn have large differences in their deformations and are strongly mixed, similar to $^{116}$Sn~\cite{PhysRevC.99.024303,Pore2017,Kantele1979}. In $^{116}$Sn, a $10^3\rho^2(E0;0^+_3\rightarrow 0^+_2)$ value of 100(20) was reported with an estimated difference in deformation parameter $\beta_2$ of at least 0.22 in Ref.~\cite{Kantele1979}, and mixing amplitudes $\alpha^2 = 0.69$ and $\beta^2 = 0.31$ were determined in Ref.~\cite{Pore2017}. If we assume, as in Ref.~\cite{Kantele1979}, that the $0^+_2$ and $0^+_3$ states are constructed from the complete mixtures of the spherical and deformed unperturbed states, the $\rho(E0)$ strength can be crudely related to the difference in deformation parameter $\Delta\beta_2$ through the matrix element $m(E0)$~\cite{Kantele1979},
\begin{equation}
    \label{eq:E0matrix}
    \rho(E0) = \cfrac{m(E0)}{eR^2} = 6.8 \Delta\langle\beta_2^2\rangle~,
\end{equation}
where $e$ is the elementary charge and $R$ is the nuclear radius. Similar to $^{116}$Sn~\cite{Kantele1979}, a minimum of $\Delta\beta_2 = 0.23(2)$~\footnote{the $\pm$2 uncertainty results from the uncertainty in $\rho^2(E0)$. However, Eq.~\ref{eq:E0matrix} is only intended for an order-of-magnitude approximation due to possible variations in the radial distributions from the Woods-Saxon potential.} is required to produce the observed $\rho^2(E0) = 0.12(5)$ in $^{120}$Sn. If the $0^+_2$ and $0^+_3$ states are less mixed, the deformation difference would need to be even larger~\cite{Kantele1979} to reproduce the $\rho^2(E0)$ value observed in this work.\par

Our measured $B(E2;0^+_3\rightarrow2^+_1)$ value of 0.50(7)~W.u. in $^{120}$Sn is almost identical to the value of 0.49(6)~W.u. in $^{116}$Sn~\cite{Bcklin1981}. This means that the increase in the $X(0^+_3\rightarrow0^+_1)$ value by a factor of 3.3 from $^{116}$Sn to $^{120}$Sn, which was observed in Ref.~\cite{Bcklin1981}, is due to the increase in the $E0$ strength, as shown in Fig.~\ref{fig:em_rates}. While both $E0$ strengths are small, the increase in the $\rho^2(E0;0^+_3\rightarrow0^+_1)$ value by a factor of 3.4 suggests that the shape difference between the $0^+_3$ and $0^+_1$ states is larger in $^{120}$Sn than in $^{116}$Sn, according to Eq.~\ref{eq:E0matrix}. While this change by itself could be a result of neutron excitations~\cite{Bcklin1981} and the related induced polarization of the nuclear core by the extra neutrons~\cite{WOOD1999323}, there is no obvious explanation to why the $\rho^2(E0;0^+_3\rightarrow0^+_1)$ triples while $\rho^2(E0;0^+_3\rightarrow0^+_2)$ remains similar from $^{116}$Sn to $^{120}$Sn. The bounds from the upper and lower limits for the transition strengths from the $0^+_3$ state in $^{118}$Sn, as shown in the middle panel of Fig.~\ref{fig:em_rates}, seems to fit into this trend, but a precise measurement is required to see the systematics more clearly. 

In summary, the newly measured lifetime of the $0^+_3$ state corresponds to $E0$ transition strengths 17 times larger than the most-recently evaluated upper limit~\cite{KIBEDI2022103930}. It suggests strong mixing and shape coexistence between the $0^+_2$ and $0^+_3$ states, and a larger shape difference between the $0^+_3$ and $0^+_1$ states in $^{120}$Sn relative to $^{116}$Sn, while the shape difference between the $0^+_3$ and $0^+_2$ states are similar in the two nuclei. Further experimental investigations are needed to elucidate whether the $0^+_2$ and $0^+_3$ states are prolate, oblate, or triaxial. We hope that this work will motivate high-statistics Coulomb excitation measurements using the state-of-art high-efficiency $\gamma$-ray arrays, which are required to firmly place these $0^+$ states on the $\beta_2$-$\gamma$ deformation plane. The $E0$ branching ratio from the $0^+_3$ state should also be measured with better precision because the 40\% uncertainty of the $\rho^2(E0;0^+_3\rightarrow 0^+_2)$ value arises almost entirely from the branching ratio. The uncertainty can be improved indirectly by high-statistics $\gamma$-ray spectroscopy through the intensity-balance technique as described in Ref.~\cite{PhysRevC.109.054317} or directly by conversion-electron spectroscopy using high-efficiency Si(Li) arrays that are sensitive around $\approx$300~keV and possibly employ a magnetic lens to suppress background by deflecting the positrons from $\beta^+$ decay~\footnote{Further coincidence conditions may be required to suppress the lower energy $\beta^-$ background that was observed in Ref.~\cite{Bcklin1981}.}. \par

The authors thank T.~Kib\'edi for valuable discussions regarding the $\rho^2(E0)$ values and their uncertainties and C.~M{\"u}eller-Gatermann for the target material. This work was supported in part by the Natural Sciences and Engineering Research Council of Canada and the U.S. Department of Energy, Office of Science, Office of Nuclear Physics, under Contracts No. DE-AC02-06CH11357~(ANL).\par 

\bibliography{120Sn_clean}

\end{document}